\documentclass{rmaa}

\title{Radio Continuum Sources Associated with AB Aur}

 \author{ Luis F. Rodr\'\i guez,\altaffilmark{1} 
          Luis Zapata,\altaffilmark{1,2}
          and Paul T. P. Ho\altaffilmark{3,2}} 
  \altaffiltext{1}{Centro de Radioastronom\'{\i}a y Astrof\'{\i}sica, Universidad
Nacional Aut\'onoma de M\'exico, Campus Morelia} 
  \altaffiltext{2}{Harvard-Smithsonian Center for Astrophysics}
  \altaffiltext{3}{Academia Sinica, Institute of Astronomy and Astrophysics}

\fulladdresses{
\item Luis F. Rodr\'\i guez and Luis Zapata: Centro de Radiostronom\'{\i}a
y Astrof\'\i sica, 
UNAM, A. P. 3-72, 
(Xangari), 58089 Morelia, Michoac\'an, M\'exico
(l.zapata; l.rodriguez@astrosmo.unam.mx).
\item Paul T. P. Ho: Academia Sinica, Institute of Astronomy and Astrophysics,
P.O. Box 23-141, Taipei 106, Taiwan and Harvard-Smithsonian Center for Astrophysics
(pho@asiaa.sinica.edu.tw), and 
Submillimeter Array, 60 Garden Street, Cambridge, MA 02138, USA
(ho@cfa.harvard.edu) 
}

\shortauthor{Rodr\'\i guez et al.}
\shorttitle{Radio continuum observations of AB Aur}

\SetVolume{43} \SetFirstPage{001} \SetYear{2007}
\ReceivedDate{2006 June 27} 
\AcceptedDate{2006 November 08}

\resumen{Presentamos observaciones de alta resoluci\'on angular y alta
sensitividad, 
hechas con el Very Large Array a 3.6 cm, de la estrella Ae de Herbig
AB Aur. Esta estrella es de inter\'es porque su disco circunestelar
muestra caracter\'\i sticas 
que se han atribuido a la presencia de una compa\~nera de baja masa o de
un planeta gaseoso gigante, a\'un sin detectar.
Nuestra imagen confirma la emisi\'on de continuo que se sabe existe
en asociaci\'on con la estrella, y detecta a una d\'ebil protuberancia
que se extiende aproximadamente $0\rlap.{''}3$ al SE de la estrella.
Consideraciones te\'oricas y resultados observacionales previos son
consistentes con la presencia de una compa\~nera de AB Aur con la separaci\'on
y \'angulo de posici\'on
derivados de nuestros datos de radio.
Tambi\'en determinamos los movimientos propios de AB Aur comparando nuestras
nuevas observaciones con datos tomados 17 a\~nos atr\'as, y encontramos
valores consistentes con los determinados por Hipparcos. 
}
 
\abstract{We present high angular resolution, high-sensitivity
Very Large Array observations at 3.6 cm of the Herbig Ae star AB Aur.
This star is of interest since its circumstellar disk
exhibits characteristics that have been attributed to the presence of
an undetected low mass companion or giant gas planet.
Our image confirms the continuum emission known to exist
in association with the star, and detects a faint protuberance that extends about
$0\rlap.{''}3$ to its SE. 
Previous theoretical considerations and observational results are consistent
with the presence of a companion to AB Aur with the separation and position
angle derived from our radio data. 
We also determine the proper motion of AB Aur by comparing our new observations
with data taken about 17 years ago and find values consistent with those
found by Hipparcos. 
}
      

\begin{document}

\maketitle

\section{Introduction}

AB Aurigae (HD 31293) is one of the nearest Herbig Ae stars. It has a spectral 
type A0-A1 (Hern\'andez et al. 2004) 
and from the Hipparcos parallax measurements (van den Ancker et al. 1998)
it is known to be located at a distance of $D=144\pm^{23}_{17}$ pc.

This star has received significant attention lately since its circumstellar disk
(Mannings \& Sargent 1997; Grady et al. 1999) was found to
exhibit complex spiral-like structures in the near-IR
continuum (Fukugawa et al. 2004), as well as in millimeter observations of molecular
lines and continuum (Corder, Eisner, \& Sargent 2005;
Pi\'etu, Guilloteau, \& Dutrey 2005; Lin et al. 2006).
Corder et al. (2005) estimate the outer radius of the disk to be
$\sim$ 600 AU.
In their CO and continuum observations of AB Aur at 3 and 1.3 mm
Pi\'etu et al. (2005) found that the disk also has non-Keplerian
motions and an inner hole about 70 AU in radius
and proposed as a possible explanation for these peculiar disk characteristics
the presence of a low mass companion located about 40 AU from AB Aur.
Lin et al. (2006) also find non-Keplerian motions in the disk and
the presence of a central depression and suggest that these 
dynamical perturbations could be produced by
the possible existence of a giant planet forming in the disk.

However, a number of optical and infrared studies
place stringent mass limits on a companion to AB Aur.
As discussed by Pi\'etu et al. (2005), these limits depend on the distance
of the companion to the star (the closer the companion is,
the harder it is to detect from imaging because of the bright stellar emission at optical
and infrared wavelengths, 
and thus the higher the upper limit). From 
the near-infrared speckle observations of Leinert et al. (1994),
Pi\'etu et al. (2005) estimate upper mass limits for a possible companion
in the range of
0.01 to 0.3 $M_\odot$ for distances between 140 to 10 AU 
($1\rlap.{''}0$ to $0\rlap.{''}07$).

In this paper we present sensitive, high angular resolution 3.6 cm
continuum observations
made with the Very
Large Array (VLA) in an attempt to search for a radio companion to AB Aur.
Stellar emission at radio wavelengths is faint, and the detection of such a 
companion would help us to understand
the peculiarities of disk. Our search for radio continuum emission from a possible companion
to AB Aur is justified since
some brown dwarfs have been observed as radio sources
(e. g. Berger et al. 2005) and it has also been speculated that giant gas
planets could be sources of detectable emission at radio wavelengths
(Farrell et al. 2004).

\section{Observations}

The observations were taken
on 2006 April 28 with
the VLA of the NRAO\footnote{The National Radio 
Astronomy Observatory is operated by Associated Universities 
Inc. under cooperative agreement with the National Science Foundation.}
at 3.6 cm in the A configuration.
We observed for a total of 10 hours with 
an effective bandwidth of 100~MHz and both circular
polarizations. The absolute amplitude calibrator was
1331+305 (with an adopted flux density of 5.21 Jy), while the
phase calibrator was
0443+346, with a bootstrapped flux density of 0.581$\pm$0.005 Jy.
The data
were edited and calibrated using the software package Astronomical Image
Processing System (AIPS) of NRAO. Cleaned maps were obtained using the
task IMAGR of AIPS and the ROBUST parameter
(Briggs 1995) of this task set to 5, to optimize sensitivity.

\section{Results}

\subsection{Main source and protuberance}

A source was detected in association with AB Aur
(see Figure 1). Three other sources were detected within a few
arcmin of AB Aur, none of them
with known counterparts at other wavelengths.
The parameters of the four sources detected are given in Table 1.
The deconvolved dimensions of the sources were obtained fitting them
with Gaussian ellipsoids using the
task IMFIT of AIPS and the fits were made taking into account
the effect of bandwidth smearing for the sources away from the phase center.
The sources VLA 1, 2, and 4 show extended dimensions at the
scale of $0\rlap.{''}3$ on one axis.
The source VLA~2 is the one associated with AB Aur and will be
discussed in detail below.
The small elongations observed in VLA 1 and 4 could indicate that they
are remote radio galaxies that will appear as elongated because of
their jet and lobe structures. However, part of the elongation could be due
to the effect of bandwidth smearing not being fully corrected by the task IMFIT.
Since VLA~2 is at the center of the field, it is not affected by
bandwidth smearing and its elongation is considered to be real.

The presence of weak centimeter radio emission in association with
AB Aur has been previously reported by G\"udel et al. (1989) and
Skinner et al. (1993).
The total flux density measured by us, 0.20$\pm$0.03 mJy, is consistent
within the noise of the observations
with the values reported previously at the same wavelength.
However, this is the first time that the image obtained is sensitive
enough to show the possible presence of structure in the radio emission.
The faint protuberance to the SE of AB Aur in the 3.6 cm image
shown in Figure 1 is most likely
real since it is at the modest but significant level of 4-$\sigma$ and since a similar structure
at the same position angle, separation, and brightness 
is not present in any of the other three sources in the field (as it would be expected
if the feature were due to some anomalous phase error effect). 
We have also checked the presence of the structure with self-calibration
made with long integration periods and the result 
make us confident that the result presented in Fig. 1 are real.

As can be seen in Figure 1, the main emission peak of the radio source
coincides within error with the Hipparcos position from Perryman et al. (1997),
corrected for the proper motion reported by these authors.
Once we subtract in the \sl (u,v) \rm plane a point source with the flux and position
of the radio peak component, we are left with a faint component displaced
about $0\rlap.{''}3$ to the SE of AB Aur (see bottom panel in Figure 1). 
The position and flux density of this possible source are given in Table 1.
We note that the flux observed by us ($\sim$70 $\mu$Jy) is much larger than the
flux density of $\sim$0.4 $\mu$Jy 
expected from a brown dwarf as LP944-20 (Berger et al. 2001), if located at the
same distance of AB Aur.

\begin{table*}[htbp]
\small
  \setlength{\tabnotewidth}{1.8\columnwidth} 
  \tablecols{6} 
  \caption{Radio Sources in the Field of AB Aur}
  \begin{center}
    \begin{tabular}{lccccc}\hline\hline
 &\multicolumn{2}{c}{Position$^a$} & Total Flux 
& Deconvolved & \\
\cline{2-3} 
VLA &  $\alpha$(J2000) & $\delta$(J2000) & Density (mJy) & 
Angular Size$^b$ & Counterpart \\ 
\hline
1 & 04 55 42.085 & +30 33 26.94 & 0.22$\pm$0.04 
& $0\rlap.{''}33 \pm 0\rlap.{''}08 \times <0\rlap.{''}2;~ 43^\circ\pm 25^\circ$ & -- \\
2 & 04 55 45.849 & +30 33 04.12 & 0.20$\pm$0.03
& $0\rlap.{''}27 \pm 0\rlap.{''}03 \times <0\rlap.{''}1;~ 139^\circ\pm 12^\circ$ & AB Aur \\
3 & 04 55 45.855 & +30 33 03.96 & 0.07$\pm$0.02
& $\sim 0\rlap.{''}2$ & Companion? \\
4 & 04 55 47.397 & +30 34 34.90 & 0.54$\pm$0.04
& $0\rlap.{''}22 \pm 0\rlap.{''}03 \times <0\rlap.{''}1;~ 19^\circ\pm 6^\circ$ & -- \\
5 & 04 55 47.541 & +30 32 00.52 & 0.13$\pm$0.03
& $<0\rlap.{''}2$ & -- \\

\hline\hline
\tabnotetext{a}{Units of right 
ascension are hours, minutes, and seconds
and units of declination are degrees, arcminutes, and arcseconds. Absolute positional accuracy
is estimated to be $0\rlap.{''}02$.}
\tabnotetext{b}{Major axis $\times$ minor axis; position angle.}
    \label{tab:1}
    \end{tabular}
  \end{center}
\end{table*}

\begin{figure*}
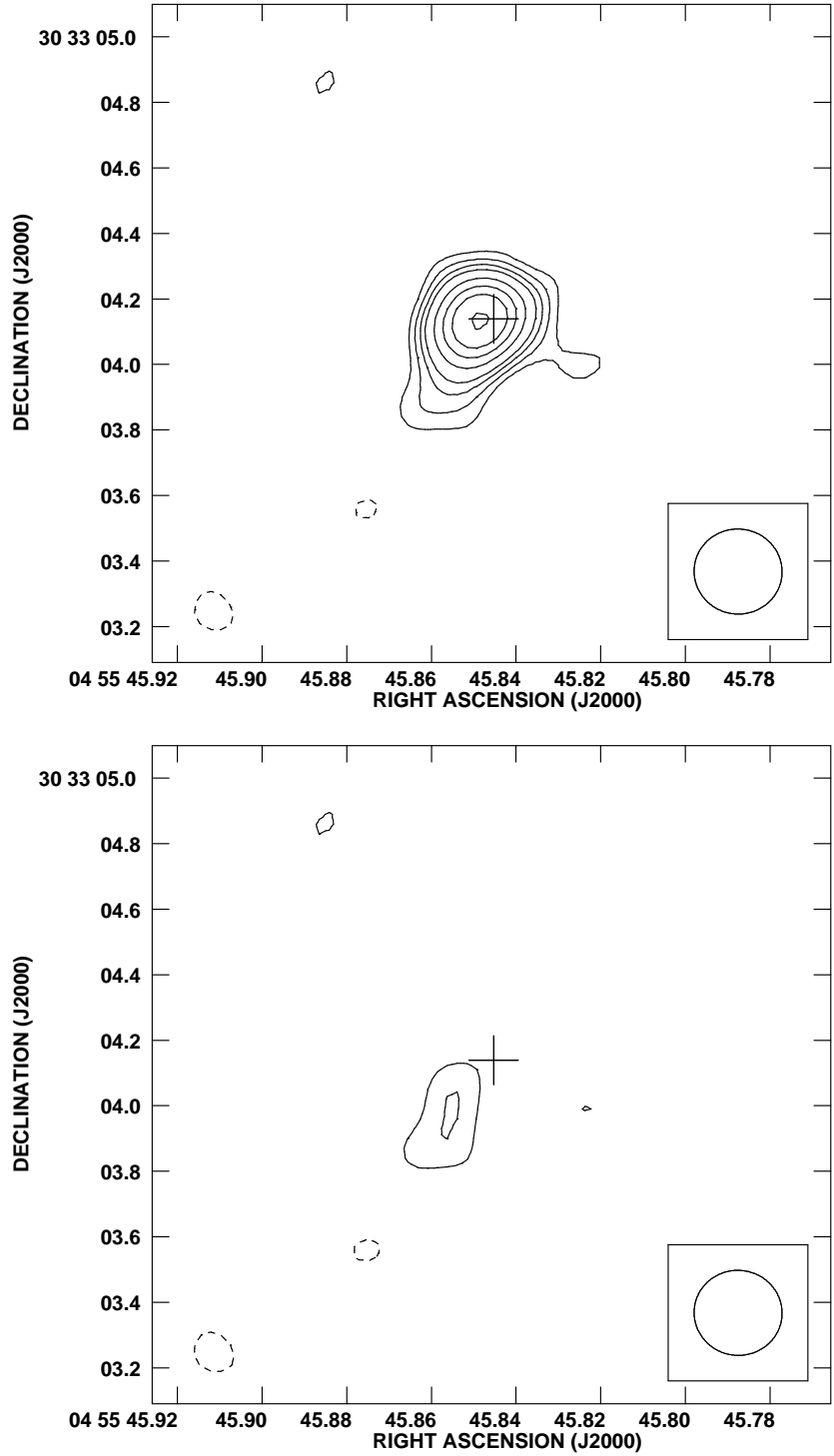

\centering
\includegraphics[scale=0.60, angle=0]{ABAURX.PS}
\includegraphics[scale=0.60, angle=0]{ABAURXMINUS.PS}
 \caption{(Top) Contour image of the 3.6 cm continuum
emission from AB Aur. (Bottom) Contour image of the 3.6 cm continuum
emission from AB Aur with a point source at the position of the
radio peak subtracted in the \sl (u,v) \rm plane. Note the residual emission in this image.
The contours are -4, -3, 3, 4, 5, 6, 8, 10, and 12 
times 9.3 $\mu$Jy, the rms noise of the image.
The cross marks the Hipparcos position of AB Aur, corrected
for proper motion. The positional error of Hipparcos for AB Aur
is $\sim0\rlap.{''}015$, taking into account the errors in absolute
position and in correction for proper motion. The size of the cross is five times
larger for clarity.
The half power contour of the beam ($0\rlap.{''}27 \times 0\rlap.{''}26;~PA ~=~+84^\circ$),
is shown in the bottom right corner.}
  \label{fig1}
\end{figure*}

\subsection{Possible explanations for the nature of the radio protuberance}

There are at least two possible explanations for the presence of the faint
protuberance in the radio image of AB Aur. The first is that we are seeing
ionized, collimated gas flowing along the position angle of $\sim$139$^\circ$. 
Thermal jets, which are ionized outflows detected at radio
wavelengths via their free-free emission (i. e. Anglada 1996; Rodr\'\i guez 1997),
are fairly common in
young stars. Furthermore, the position angle of the feature ($139^\circ \pm 12^\circ$)
is close to being perpendicular to the position angle of the major axis of the disk
($\sim$66$^\circ$), as determined from the millimeter observations of
Lin et al. (2006). This relative orientation is expected in models in which the
magnetic fields of the disk help accelerate and collimate the jet.
On the other hand, there is no evidence of optical jets
(Grady et al. 1999) or molecular outflows (Cant\'o et al. 1983;
Levreault 1988) in association with AB Aur. This may be due to the fact
that AB Aur is a star for which 
DeWarf et al. (2003) estimate an age in the range of 1 to 4 million years.
By this age, it is believed that strong outflow activity is no
longer present in young stars (Mundt, Brugel, \& B\"uhrke 1987) and
we consider this explanation unlikely.

A second possibility is that we are observing a faint radio companion to
AB Aur. In this case, the emission would most probably be
of gyrosynchrotron nature, as observed in young stars with active
magnetospheres (G\"udel 2002). This hypothetical companion would be located at about $0\rlap.{''}3$
from AB Aur, that corresponds to a distance of about 40 AU. Remarkably, 
this is the separation proposed by Pi\'etu et al. (2005) on dynamical 
considerations based on the size of the central hole in the disk. Optical and 
infrared searches with high angular resolution and sensitivity 
at this position could be worthwhile.

Another interesting result that supports the presence of a companion at the
position of the radio protuberance is that presented by
Baines et al. (2006). These authors did high-resolution optical spectro-astrometry
of AB Aur and concluded that it has a companion located at a position angle of
$146^\circ$, a value very close to that determined by us for the radio
protuberance ($139^\circ$). Baines et al. (2006) also set a lower limit to the
separation of $0\rlap.{''}026$, consistent with the separation measured by us.  
However, it should be noted that Baines et al. (2006) favor a value of
$0\rlap.{''}5$ for the separation and that this value would be inconsistent with
out results.

Unfortunately, it is difficult to be sure of the reality of the protuberance.
This structure is very faint and is below the sensitivity of previous images.
Observations at several wavelengths with high sensitivity are required but this
is not feasible now, at least within reasonable integration times. 
We will have to wait for several years to
confirm or refute the presence of a faint radio companion
to AB Aur, once the
more sensitive Expanded Very Large Array is completed. 

\subsection{Radio Proper Motions}

Finally, we use our new position and that obtained from a reanalysis of the
available VLA archive data in the A configuration for 1988 October 7 and 27 and 1990
February 10 and 12 to search for proper motions in the radio source
associated with AB Aur.
Our observations correspond to the epoch 2006.32, while those of the 
archive data are taken to have an average epoch of 1989.46.
The position of the source for this earlier epoch is 
$\alpha(2000) =  04^h~ 55^m~ 45\rlap.^s839;~ \delta(2000) = 30^\circ~ 33'~ 04\rlap.{''}59$.
Comparing with the position given in Table 1 and taken into account that
the time difference between observations is of 16.86 years, we derive
proper motions of $\mu_\alpha = +8 \pm 5~ mas~yr^{-1}$ and
$\mu_\delta = -28 \pm 5~ mas~ yr^{-1}$. These values are in 
agreement (although they have significantly less signal-to-noise ratio)
with those reported by Hipparcos (Perryman et al. 1997):
$\mu_\alpha = +1.71 \pm 1.06~ mas~yr^{-1}$ and
$\mu_\delta = -24.24 \pm 0.67~ mas~ yr^{-1}$. 
The near coincidence between the radio and optical positions shown in Figure 1 
also indicates the agreement of the radio and optical astrometries. 

Two of the remaining sources listed in
Table 1 were detected in both the 1989.46 and 2006.32 images.
The first source is VLA~1, that for the 1989.46 epoch has
a position of $\alpha(2000) =  04^h~ 55^m~ 42\rlap.^s088;~ \delta(2000) =
30^\circ~ 33'~ 26\rlap.{''}85$,
about $0\rlap.{'}9$ to the west of AB Aur.
Comparing with the position given in Table 1, we derive
proper motions of $\mu_\alpha = -2 \pm 3~ mas~yr^{-1}$ and
$\mu_\delta = +5 \pm 3~ mas~ yr^{-1}$.
The other source is VLA 3, that for the 1989.46 epoch has
a position of $\alpha(2000) =  04^h~ 55^m~ 47\rlap.^s398;~ \delta(2000) = 
30^\circ~ 34'~ 34\rlap.{''}91$,
about $1\rlap.{'}5$ to the north of AB Aur.
Again, comparing with the position given in Table 1, we derive
proper motions of $\mu_\alpha = -1 \pm 2~ mas~yr^{-1}$ and
$\mu_\delta = -1 \pm 2~ mas~ yr^{-1}$. 
Thus, in constrast to the radio source associated with AB Aur that
shows clear proper motions, the measurements of
VLA 1 and 4 are consistent with no significant proper
motions, which supports the possibility that they are extragalactic sources.

\section{Conclusions}

Our main conclusions are as follows.

1) We obtained a high angular resolution, high sensitivity 3.6 cm image
of AB Aur. Besides the emission associated with the star, that
was previously known, we detect a faint protuberance to the SE of the
star that could be tracing either a collimated outflow or possibly a
low mass companion or even a giant gas planet. In the case of a companion,
the separation measured by us ($0\rlap.{''}3$) is consistent with
that proposed by Pi\'etu et al. (2005) from 
dynamical considerations on the size of the central hole in the disk.
Furthermore, the position angle measured by us for the possible
companion ($139^\circ$) is very similar to that determined by
Baines et al. (2006) from high-resolution optical spectro-astrometry
of AB Aur.
However, the emission is
faint and requires confirmation with future, more sensitive facilities.

2) Comparing with data taken about 17 years before, we can determine
in the radio wavevelengths the proper motion 
of AB Aur, that is consistent with that measured with Hipparcos.
In contrast, two other sources in the surroundings (VLA 1 and VLA 4) show no
detectable proper motions, suggesting they are background extragalactic sources.


\acknowledgments
LFR acknowledges the support
of DGAPA, UNAM, and of CONACyT (M\'exico).
This research has made use of the SIMBAD database, 
operated at CDS, Strasbourg, France.


\end{document}